\documentclass[twocolumn,prb]{revtex4}

\usepackage{color}
\usepackage{array}
\usepackage{amsmath}
\usepackage{amssymb}
\usepackage{wasysym}
\usepackage{bm}
\usepackage{graphicx}
\usepackage{txfonts}
\usepackage{epsfig}

\begin{document}

\title{General procedure for determining braiding and statistics of anyons using entanglement interferometry}

\author{Yi Zhang}
\affiliation{Department of Physics, Stanford University, Stanford, California 94305, USA}

\author{Tarun Grover}
\affiliation{Kavli Institute for Theoretical Physics, University of California, Santa Barbara, CA 93106, USA}

\author{Ashvin Vishwanath}
\affiliation{Department of Physics, University of California,
Berkeley, CA 94720, USA}

\begin{abstract}
Recently, it was argued that the braiding and statistics of anyons
in a two-dimensional topological phase can be extracted by studying
the quantum entanglement of the degenerate ground-states on the
torus. This construction either required a lattice symmetry (such as
$\pi/2$ rotation) or tacitly assumed that the `minimum entanglement
states' (MESs) for two different bipartitions can be uniquely
assigned quasiparticle labels. Here we describe a procedure to
obtain the modular $\mathcal S$ matrix, which encodes the braiding
statistics of anyons, which does not require making any of these
assumptions. Our strategy is to coherently compare MESs of three
independent entanglement bipartitions of the torus, which leads to a
unique modular $\mathcal S$. This procedure also puts strong
constraints on the modular $\mathcal T$ and $\mathcal U$ matrices
without requiring any symmetries, and in certain special cases,
completely determines it. Our method applies equally to Abelian and
non-Abelian topological phases.
\end{abstract}

\maketitle

\section{Introduction}

Topological ordered phases in two dimensions such as fractional
quantum Hall states and quantum spin-liquids are characterized by
the presence of anyonic excitations which satisfy specific braid
statistics rules when taken around each other \cite{wen2004,
anderson1987, wen1990, Read89, wen1991, read1991, senthil2000,
sondhi,PatrickLeeReview,BalentsReview}. The presence of anyons
implies that a topological ordered phase possesses a degenerate set
of ground states on a torus \cite{wu1991, tao1984, wen_niu1990,
wen_fradkin1990, senthil2005}. The intimate relation between the
ground-state degeneracy and the presence of anyons suggests that the
braid statistics rules must be encoded in the degenerate ground
states themselves. In this paper, we generalize the discussion in
Ref. \onlinecite{zhang2012} to obtain the braiding and statistics
from the ground states.

Mathematically, the braiding statistics is encoded in the modular
$\mathcal{S}$ and $\mathcal{U}$ matrices (or equivalently,
$\mathcal{S}$ and $\mathcal{T}$ matrices where $\mathcal{U} =
\mathcal{T} \mathcal{S} \mathcal{T}$)
\cite{wen1990,wen93,kitaev2006,yellowbook}. The $\mathcal{S}$ matrix
expresses the mutual statistics between anyons while the
$\mathcal{U}$ matrix encodes the self-statistics. For chiral
topological phases, there is an additional parameter, the central
charge $c$ for the edge states, which is determined modulo $8$ by
the $\mathcal{S}$ and $\mathcal{U}$ matrices \cite{kitaev2006}. Ref.
\onlinecite{zhang2012} argued that the modular $\mathcal{S}$ and
$\mathcal{U}$ matrices can be determined by calculating the quantum
entanglement of degenerate ground states. This builds on the idea
that the scaling of entanglement entropy with respect to the
subsystem size often yields universal information about the
corresponding phase of matter. For example, the von Neumann
entanglement entropy $S_{vN}$ of a gapped ground state for a
contractible disk-shaped region in two dimensions with a boundary of
size $\ell$ is given by $S_{vN} = \alpha \ell - \gamma + O(1/\ell)$,
where $\gamma$ is the so-called `topological entanglement
entropy'(TEE) and is given by $\gamma = \sqrt{\sum d^2_a}$ with
$d_a/\gamma$ being the first row of the modular $\mathcal{S}$ matrix
\cite{hamma2005, kitaev2006, levin2006}.

This result motivates one to ask whether the full modular
$\mathcal{S}$ and $\mathcal{U}$ matrices might also be extractable
from the ground-state entanglement. Ref. \onlinecite{zhang2012}
argued that the answer is indeed positive. The basic idea mainly
consists of two steps: first, given a set of degenerate ground
states $\left|\xi_{a}\right\rangle $, $a=1, \ldots, N$, the TEE
corresponding to a generic ground state $|\psi\rangle = \sum c_a
|\xi_{a}\rangle$ for \textit{non-contractible} subregions on a
torus, e.g. partitioning the torus into two cylinders, generally
differs from the value in trivial subregions, and is maximized by a
special set of coefficients $c_a$, which can be identified using TEE
as an indicator. Note that the TEE reduces the total entanglement
entropy, such a state minimizes the total entanglement entropy and
is therefore dubbed as the `minimum entropy state' (MES). Given a
non-contractible entanglement bipartition, the complete set of MESs
forms a basis of the degenerate ground states, which correspond to
the simultaneous eigenstates of the Hamiltonian as well as the
operators that measure the quasiparticles through the boundary cycle
of the entanglement bipartition. We hereafter denote the MESs basis
for a bipartition $\alpha$ as $\left\{
\left|\Xi^{(\alpha)}_a\right\rangle\right\}\equiv\left|\Xi^{(\alpha)}\right\rangle$.
Then, the modular matrices can be related to the unitary
transformations between two inequivalent sets of MESs defined for
two different bipartitions ($\alpha$'s). For example, having
identified the eigenstates of operators that measure quasiparticles,
viz. the MES states, the elements of the modular $\mathcal{S}$
matrix are given by the overlap: $\mathcal S_{ab} = \left\langle
\Xi^{(1)}_a|\Xi^{(2)}_b\right\rangle$ where the two entanglement
bipartitions are along the $\hat{x}$ and $\hat{y}$ directions,
respectively. This procedure can be further simplified in the
presence of certain rotation symmetry $R$, which relates the two
sets MESs
$\left|\Xi^{(2)}\right\rangle=R\left|\Xi^{(1)}\right\rangle$. For
example, $R_{\pi/2}:\hat x\rightarrow \hat y$ gives $\mathcal
S=\left\langle\Xi^{(1)}\right|
R_{\pi/2}\left|\Xi^{(1)}\right\rangle$.

This method was successfully incorporated into matrix product state
(MPS) and DMRG based techniques for finding ground states in
quasi-2D systems \cite{cincio2013, zaletel2013, zhu2013}, as well as
new applications via variational Monte Carlo \cite{zhu2013,
jiang2014}. A slightly different method to obtain modular matrices
was recently proposed in several papers\cite{wen2014}. It is also
worth noting that for chiral topological phases, momentum
polarization method \cite{tu2013, zhang2014} can also be used to
obtain partial topological data, viz., self-statistics of anyons and
the chiral central charge. Finally, it was shown \cite{haah2014}
that for a restricted class of Hamiltonians that can be written as
sum of local commuting projectors, under certain reasonable
assumptions one can obtain the modular matrix $\mathcal{S}$ using a
single ground state.

An important detail in calculating an inner product such as
$\left\langle \Xi^{(1)}_a|\Xi^{(2)}_b\right\rangle$, which was
presumed in Ref. \onlinecite{zhang2012}, is the relative ordering of
the sets of MESs $\left\{\left|\Xi^{(1)}_a\right\rangle\right\}$ and
$\left\{\left|\Xi^{(2)}_a\right\rangle\right\}$. In the presence of
a consistent point group symmetry of the lattice, the relative
ordering is automatically fixed\cite{zhang2012}. This is however not
true when such a symmetry is absent. This statement also holds true
for extracting the modular $\mathcal{U}$ matrix where one utilizes
overlap of MESs corresponding to bipartitions that differ by an
angle of $2\pi/3$. An additional related concern is to identify the
MES corresponding to the `identity quasiparticle': conventionally,
the first row and column of the $\mathcal{S}$ and $\mathcal{U}$
matrices is labeled by the identity quasiparticle. However, it is
not obvious how the method described in Ref. \onlinecite{zhang2012}
makes such an identification.

In this paper, we show that a \textit{third} entanglement
bipartition (in addition to the two bipartitions used in Ref.
\onlinecite{zhang2012}) helps to resolve this ambiguity. By
considering the `entanglement interferometry' that consists of a
series of modular transformations that start and end with the same
set of MESs, one can effectively cancel out the impact of the
unknown details of the intermediate MESs. This procedure completely
fixes the relative ordering of MESs for the disparate bipartitions.
This also fixes the identification of quasiparticles vis-a-vis the
MES states \textit{upto an Abelian quasiparticle}. Our main result
is that the modular $\mathcal{S}$ matrix is uniquely determined by
considering overlap of MESs obtained from three entanglement
bipartitions. As far as the modular $\mathcal{U}$ matrix is
concerned, in the absence of any symmetries we are able to determine
it only upto an additional phase factor for each quasiparticle,
which is further constrained by the modular $\mathcal{S}$ matrix. In
special cases, the constraints are strong enough to determine the
$\mathcal{U}$ matrix fully, without requiring any symmetry.

The rest of the paper is organized as follows: In Sec.
\ref{sec:mes}, we briefly review the method in Ref.
\onlinecite{zhang2012} and discuss its shortcoming in the absence of
spatial symmetry; to resolve this issue, we propose in Sec.
\ref{sec:smat} a general algorithm with a third entanglement
bipartition to extract the modular $\mathcal S$ matrix; in Sec.
\ref{sec:umat}, we study its further application on the modular
$\mathcal U$ matrix and quasiparticle spin of the topological
ordered state; three illustrative examples are discussed in Sec.
\ref{sec:example}.

\section{The minimum entropy states and modular
matrices}\label{sec:mes}

For concreteness, let us denote $\left|\xi_{a}\right\rangle $, $a=1,
\ldots, N$ as the complete, orthonormal set of the degenerate ground
states that we will use as our basis for the entire ground-state
manifold.

For a given nontrivial entanglement bipartition, the MESs are by
definition the ground states with minimum entanglement entropy
(maximum TEE) and can be generated by $T^{(\alpha)}_p$: the
insertion of the quasiparticles of the topological ordered state
labeled by $p$ through the non-contractible cycle enclosed by the
$\alpha$'s entanglement bipartition boundary\cite{zhang2012}. These
MESs can be sequentially obtained by maximizing the TEE in the
parameter space of the entire ground-state manifold
$\underset{a}{\sum}c_a \left|\xi_{a}\right\rangle$. Let us denote
the resulting MESs as
\begin{eqnarray} \left|\Xi_{b}^{(\alpha)}\right\rangle
=e^{i\phi_{b}^{(\alpha)}}U^{(\alpha)}_{ab}\left|\xi_{a}\right\rangle
\label{eqn:udef}
\end{eqnarray}
where the superscript $\alpha$ labels entanglement bipartitions with
inequivalent non-contractible boundaries. $\phi_{b}^{(\alpha)}$ is
an undetermined phase factor for each MES, which does not affect the
resulting entanglement entropy.

The transformation between two MES bases is essentially a modular
transformation $\mathcal{F(S,U)}$ \cite{zhang2012}. Such a
transformation can be also viewed as the transformation of the
primitive vectors that define the torus and encoded in the
$\ensuremath{SL(2,\mathbb{Z})}$ modular matrix $F(S,U)$, which can
be expressed in terms of the two generators of $SL(2,\mathbb{Z})$
\begin{eqnarray}
S&=&\left(\begin{array}{cc}
0 & 1\\
-1 & 0
\end{array}\right)\nonumber \\
U&=&\left(\begin{array}{cc}
1 & 1\\
0 & 1
\end{array}\right)\label{eqn:sl2gen}\end{eqnarray}

\begin{figure}
\begin{center}
\includegraphics[scale=0.35]{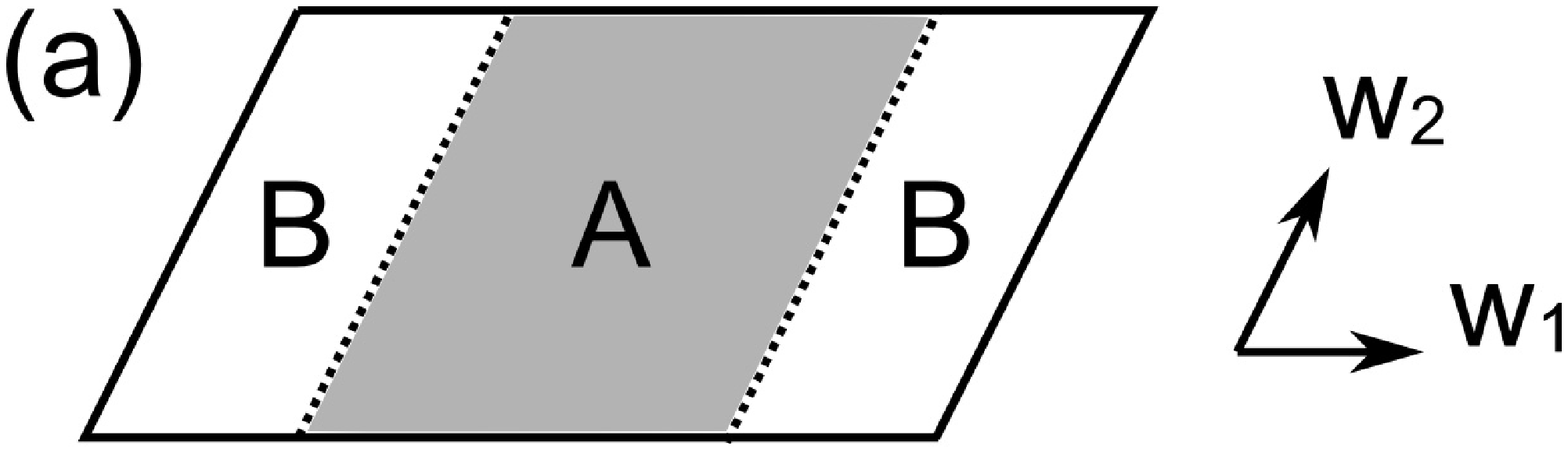}
\includegraphics[scale=0.35]{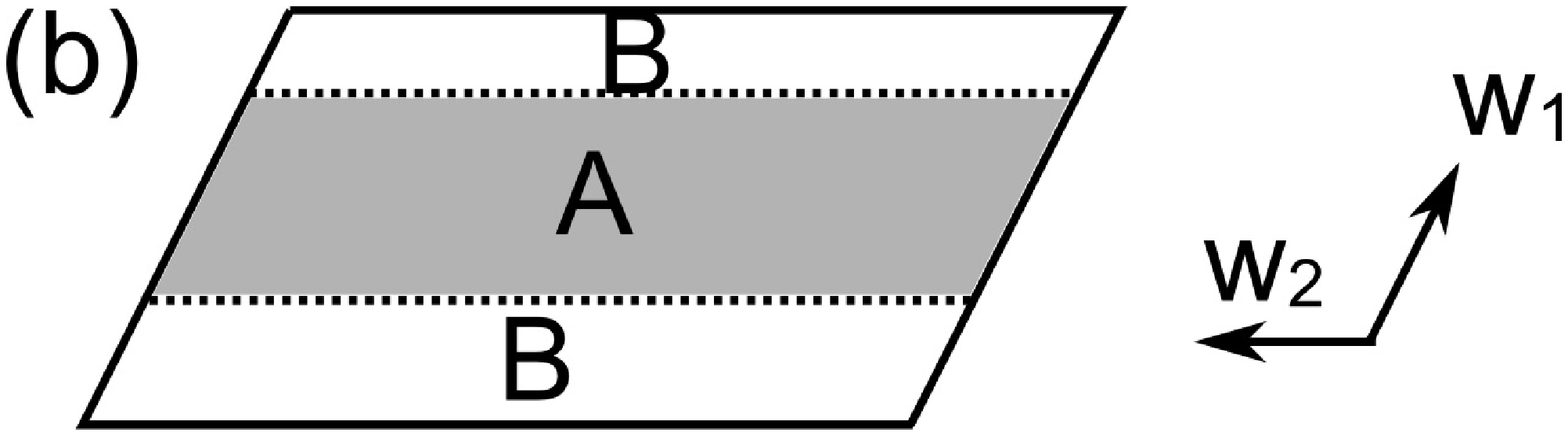}
\caption{Left panel: two nontrivial entanglement bipartitions shown
as the dashed lines that separate the system into two cylindrical
subsystems (periodic boundary condition is assumed through all
boundaries); Right panel: two sets of primitive vectors for the same
torus, we use $\vec{w}_{2}$ to label the boundary direction of each
entanglement bipartition.} \label{fig1}
\end{center}
\end{figure}

For concreteness, let us consider a modular transformation from the
primitive vectors $\vec{w}_{i=1,2}^{(1)}$ in Fig. \ref{fig1}(a) to
$\vec{w}_{i=1,2}^{(2)}$ in Fig. \ref{fig1}(b), where we use
$\vec{w}_{2}^{(\alpha)}$ to label the direction along the
non-contractible boundary of the $\alpha$th entanglement
bipartition. The relation between the two sets of primitive vectors
is
\begin{eqnarray}
\vec{w}_{1}^{(2)}&=&\vec{w}_{2}^{(1)}
\nonumber \\
\vec{w}_{2}^{(2)}&=&-\vec{w}_{1}^{(1)}\label{eqn:w1w2}
\end{eqnarray}

According to Eqn. \ref{eqn:sl2gen}, this implies $F(S,U)=S$, and
therefore the transformation between the two MES bases
$\left\langle\Xi^{(2)}|\Xi^{(1)}\right\rangle=\left(V^{(2)}\right)^{-1}\left({U}^{(2)}\right)^{-1}U^{(1)}V^{(1)}$
is equivalent to the modular $\mathcal{S}$ matrix, where we denote
the diagonal phase factors in Eqn. \ref{eqn:udef}
$V^{(\alpha)}=\mbox{diag}(e^{i\phi_{b}^{(\alpha)}})$. Since the
undetermined $V^{(\alpha)}$ are diagonal phase factors, they are
straightforward to determine with the knowledge that the elements of
the first column and row of the modular $\mathcal{S}$ matrix are
real and positive in accord with the definition of the identity
particle\cite{zhang2012}
\begin{eqnarray}
\mathcal{S}=\left(V^{(2)}\right)^{-1}\left({U}^{(2)}\right)^{-1}U^{(1)}V^{(1)}=R\left[\left({U}^{(2)}\right)^{-1}U^{(1)}\right]
\label{eqn:ou2u1}
\end{eqnarray}
where the function $R[X]$ corresponds to left and right matrix multiplication of the matrix $X$ with certain diagonal phase
factors, respectively, so as to make the elements of the first
column and row of $X$ real and positive.

However, there is an important complication that we have overlooked:
in the derivation of Eqn. \ref{eqn:ou2u1}, we have assumed the same
ordering for the two sets of MESs, while the maximum TEE requirement
makes no distinction between the MESs with the same quantum
dimension\cite{zhang2012, zhang2013}, in particular the MESs
connected with the Abelian quasiparticles (including the identity
particle) which have quantum dimension $1$. Therefore, the ordering
and particle content of the obtained MESs remain largely
undetermined. In the absence of symmetry, the orderings of the MESs
for different entanglement bipartitions are generally unrelated and
may result in a scramble of the rows and the columns of the
resulting modular matrices.

To be more specific, we now need to generalize Eqn. \ref{eqn:udef}
by including possible permutations within each set of the MESs
\begin{eqnarray}
U^{(\alpha)}_{ab}&=&\bar U^{(\alpha)}_{ab'} P^{(\alpha)}_{b'b} \nonumber\\
\left|\Xi_{b}^{(\alpha)}\right\rangle &=&\bar
U^{(\alpha)}_{ab'}V^{(\alpha)}_{b'b'}
P^{(\alpha)}_{b'b}\left|\xi_{a}\right\rangle
\end{eqnarray}
where we have introduced a permutation matrix $P_{\alpha}$ that
permutes the columns of $U^{(\alpha)}$. As we have argued above, TEE
calculations only determine $\bar U^{(\alpha)}$.

Following previous argument, we obtain instead of Eqn.
\ref{eqn:ou2u1}
\begin{eqnarray}
\mathcal{S}&=&\left(P^{(2)}\right)^{-1}\left(V^{(2)}\right)^{-1}\left(\bar
U^{(2)}\right)^{-1}\bar U^{(1)}V^{(1)}P^{(1)}
\nonumber \\
P^{(2)}\mathcal{S}\left(P^{(1)}\right)^{-1}&=&\left(V^{(2)}\right)^{-1}\left(\bar
U^{(2)}\right)^{-1}\bar U^{(1)}V^{(1)} \label{eqn:u2u1}
\end{eqnarray}

Due to the presence of the undetermined $P^{(1)}$ and $P^{(2)}$, we
can no longer assume the elements of the first column and row of
$\left(\bar U^{(2)}\right)^{-1}\bar U^{(1)}$ to be real and
positive. In particular, the
 matrix $R\left[\left(\bar U^{(2)}\right)^{-1}\bar
U^{(1)}\right]$ does not give the correct modular $\mathcal S$
matrix in general (symmetry of $\mathcal{S}$ matrix turns out to be
insufficient to uniquely pick out the correct answer consistent with
Eqn.\ref{eqn:u2u1}).

\section{General algorithm for the modular $\mathcal S$ matrix}
\label{sec:smat}

\begin{figure}
\begin{center}
\includegraphics[scale=0.35]{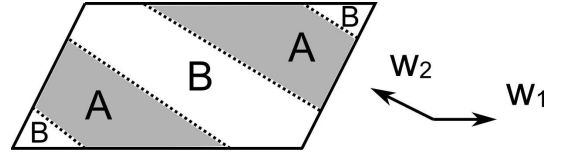}
\caption{Left panel: an additional and inequivalent entanglement
bipartition scheme; Right panel: the corresponding primitive
vectors, where $\vec{w}_{2}$ labeling the boundary direction of the
entanglement bipartition is the sum of the previous two in Fig.
\ref{fig1}.} \label{fig2}
\end{center}
\end{figure}

To resolve the difficulty of properly ordering MESs, we now present
a general algorithm to extract the modular $\mathcal S$ matrix for a
generic topological ordered state with no implicit spatial
symmetries. We find it fruitful to introduce an additional
entanglement bipartition along the $\vec w_2^{(3)} = \vec w_2^{(1)}
+ \vec w_2^{(2)}$ direction and derive the corresponding MESs
$\left|\Xi^{(3)}\right\rangle$, see Fig. \ref{fig2} for
illustration. Without loss of generality, we assume that for all
three entanglement bipartitions, the first MES is associated with an
Abelian quasiparticle, which in practice can be verified by making
sure that the corresponding quantum dimension, as obtained via TEE,
is $1$. Then, as we shall prove below, the modular $\mathcal S$
matrix is fully determined through
\begin{eqnarray}
\mathcal{S} &=& \left\{R\left[\left(\bar U^{(2)}\right)^{-1}\bar
U^{(1)}\right]\right\}^{-1}R\left[\left(\bar U^{(2)}\right)^{-1}\bar
U^{(3)}\right]R\left[\left(\bar U^{(3)}\right)^{-1}\bar
U^{(1)}\right]\nonumber\\
\label{eqn:modS}
\end{eqnarray}
this is the first main conclusion of this paper. Again, the function
$R[X]$ corresponds to left and right matrix multiplication of the matrix $X$ with certain diagonal phase
factors, respectively, so as to make the elements of the first
column and row of $X$ real and positive.

To derive Eqn.\ref{eqn:modS}, we first note that the primitive
vectors of the third entanglement bipartition may also be expanded
as
\begin{eqnarray}
\vec{w}_{1}^{(3)}&=&\vec{w}_{1}^{(1)}
\nonumber \\
\vec{w}_{2}^{(3)}&=&-\vec{w}_{1}^{(1)}+\vec{w}_{2}^{(1)}
\label{eqn:w2w3}
\end{eqnarray}
and equally
\begin{eqnarray}
\vec{w}_{1}^{(2)}&=&\vec{w}_{1}^{(3)}+\vec{w}_{2}^{(3)}
\nonumber \\
\vec{w}_{2}^{(2)}&=&-\vec{w}_{1}^{(3)} \label{eqn:w1w3}
\end{eqnarray}
In the Appendix, we further discuss the benefits of choosing  $\vec w_2^{(3)} = \vec
w_2^{(1)} + \vec w_2^{(2)}$ for the third entanglement bipartition.

With  arguments similar to the last section, the modular matrices
corresponding to the transformations from
$\left|\Xi^{(1)}\right\rangle$ to $\left|\Xi^{(3)}\right\rangle$ and
from $\left|\Xi^{(3)}\right\rangle$ to
$\left|\Xi^{(2)}\right\rangle$ are $\mathcal{U}^{-1}\mathcal
S\mathcal{U}^{-1}$ and $\mathcal{U}^{-1}\mathcal {S}$, respectively.
Therefore, concerning the third entanglement bipartition we have
\begin{eqnarray}
\mathcal{U}^{-1}\mathcal
S\mathcal{U}^{-1}&=&\left(P^{(3)}\right)^{-1}\left(V^{(3)}\right)^{-1}\left(\bar
U^{(3)}\right)^{-1}\bar U^{(1)}V^{(1)}P^{(1)}
\nonumber \\
\mathcal{U}^{-1}\mathcal
{S}&=&\left(P^{(2)}\right)^{-1}\left(V^{(2)}\right)^{-1}\left(\bar
U^{(2)}\right)^{-1}\bar U^{(3)}V^{(3)}P^{(3)} \label{eqn:u3u1}
\end{eqnarray}

Next, without loss of generality, it is straightforward to separate
each permutation $P^{(\alpha)}$ into two distinct parts:
$P^{(\alpha)}=\bar P^{(\alpha)}\tilde P^{(\alpha)}$, where $\bar
P^{(\alpha)}$ is the permutation on columns other than the first
one, while $\tilde P^{(\alpha)}$ maps each state to that with an
\emph{additional} Abelian quasiparticle, as is determined by the
particle content of the first MES. By definition, any MES can be
obtained from the MES associated with the identity quasiparticle
through quasiparticle insertion
\begin{eqnarray}
\mathcal{S}_{ab}=\left\langle\Xi^{(2)}_a|\Xi^{(1)}_b\right\rangle
\equiv \left\langle 1^{(2)}\left| [T_a^{(2)}]^{-1}|T^{(1)}_b\right|
1^{(1)}\right\rangle
\end{eqnarray}
where $\left| 1^{(\alpha)}\right\rangle$ and $T^{(\alpha)}_a$ are
the MES associated with the identity particle and insertion operator
of an quasiparticle $a$ along the $\vec w^{(\alpha)}_2$ direction,
respectively. In comparison,
\begin{eqnarray}
\mathcal{S}_{ab}\left(\tilde
P^{(1)}\right)^{-1}&=&\left\langle\Xi^{(2)}_a\left|T^{(1)}_{p_1}
\right|\Xi^{(1)}_b\right\rangle \nonumber \\&=& \left\langle
1^{(2)}\left|
[T_a^{(2)}]^{-1}|T^{(1)}_{p_1} T^{(1)}_b\right| 1^{(1)}\right\rangle\nonumber\\
&=&\left\langle 1^{(2)}\left|T^{(1)}_{p_1} [T_a^{(2)}]^{-1}|
T^{(1)}_b\right| 1^{(1)}\right\rangle \theta_{{p_1}\times a}/\theta_{{p_1}}\theta_{a} \nonumber \\
&=&\left\langle 1^{(2)}\left| [T_a^{(2)}]^{-1}|
T^{(1)}_b\right| 1^{(1)}\right\rangle \theta_{{p_1}\times a}/\theta_{{p_1}}\theta_{a}\nonumber \\
&=& \left(\theta_{{p_1}\times a}/\theta_{{p_1}}\theta_{a}\right)
\cdot \mathcal{S}_{ab}
\end{eqnarray}
where $\times$ is the fusion product, and $p_1$ is the added Abelian
quasiparticle by $P^{(1)}$ with topological spin $\theta_{p_1}$.
$\theta_{{p_1}\times a}/\theta_{{p_1}}\theta_{a}=[T_a^{(2)}]^{-1}
T^{(1)}_{p_1} T^{(2)}_a [T^{(1)}_{p_1}]^{-1}$ is the braiding
between quasiparticles $a$ and ${p_1}$. Physically, it is the
induced phase for adiabatically moving $a$ around ${p_1}$ and also
encoded as the phase of the matrix element $\mathcal S_{{p_1} a}$.
We have also used the fact that for an Abelian quasiparticle ${p_1}$
\begin{eqnarray}\left\langle 1^{(2)}\right|
T^{(1)}_{p_1}&=&\underset{b}{\sum} \mathcal S_{1 b} \left\langle
b^{(1)}\right| T^{(1)}_{p_1} \nonumber= \underset{b}{\sum} \mathcal
S_{1 b} \left\langle \left(b\times {p_1}\right)^{(1)}\right| \\&=&
\underset{b}{\sum} \mathcal S_{1 b\times {p_1}} \left\langle
\left(b\times {p_1}\right)^{(1)}\right| = \left\langle
1^{(2)}\right|
\end{eqnarray}
More generally, denoting the added Abelian quasiparticle by
$P^{(2)}$ as $p_2$, one can show that
\begin{eqnarray}
\tilde P^{(2)}\mathcal{S}_{ab}\left(\tilde
P^{(1)}\right)^{-1}&=&\left\langle\Xi^{(2)}_a\left|[T^{(2)}_{p_2}]^{-1}
T^{(1)}_{p_1}\right|\Xi^{(1)}_b\right\rangle \nonumber\\&=&
\left\langle 1^{(2)}\left|
[T^{(2)}_a]^{-1} [T^{(2)}_{p_2}]^{-1} | T^{(1)}_{p_1} T^{(1)}_b\right| 1^{(1)}\right\rangle\nonumber\\
&=&\left(\theta_{{p_1}\times {p_2}}/\theta_{p_1}\theta_{p_2}\right)
\left(\theta_{{p_1}\times a}/\theta_{p_1}\theta_{a}\right)\cdot
\nonumber
\\ & & \mathcal{S}_{ab} \cdot \left(\theta_{b\times
{p_2}}/\theta_{b}\theta_{p_2}\right)\nonumber\\ \label{eqn:tst}
\end{eqnarray}

Heuristically, the $ij^{th}$ component of the modular $\mathcal S$
matrix encodes the braiding between the $i^{th}$ and $j^{th}$
quasiparticles. Eqn.\ref{eqn:tst} formalizes this intuition so that
the insertion of additional Abelian quasiparticles $p_1$ and $p_2$
leads to additional Abelian phases induced by the braiding between
the $i^{th}$ and $p_1$, $p_2$ and the $j^{th}$ as well as $p_1$ and
$p_2$ quasiparticles. To this end, the additional quasiparticles
inserted by $\tilde P^{(\alpha)}$ and $\left(\tilde
P^{(\beta)}\right)^{-1}$ effectively contribute additional phase
factors to each rows and columns of the modular $\mathcal S$ matrix.

Now we note that (1) the undetermined components including
$V^{(\alpha)}$, $\mathcal U$ and $\theta_{p_1 \times a}/\theta_{p_1}
\theta_a$, etc. are all diagonal phase factors, and (2) $\bar
P^{(\alpha)}$ ($\left(\bar P^{(\beta)}\right)^{-1}$) only involves
the rows (columns) other than the first one, thus the elements in
the first line and column of $\bar P^{(\alpha)} \mathcal{S}
\left(\bar P^{(\beta)}\right)^{-1}$ remain real and positive just as
in $\mathcal S$. Therefore, Eqns. \ref{eqn:u2u1}, \ref{eqn:u3u1} and
\ref{eqn:tst} together imply
\begin{eqnarray}
\bar P^{(2)}\mathcal{S} \left( \bar P^{(1)}\right)^{-1}&=& R\left[\left(\bar U^{(2)}\right)^{-1}\bar U^{(1)}\right]\nonumber\\
\bar P^{(2)}\mathcal{S} \left( \bar P^{(3)}\right)^{-1}&=& R\left[\left(\bar U^{(2)}\right)^{-1}\bar U^{(3)}\right]\nonumber\\
\bar P^{(3)}\mathcal{S} \left( \bar P^{(1)}\right)^{-1}&=&
R\left[\left(\bar U^{(3)}\right)^{-1}\bar U^{(1)}\right]
\label{eqn:psp}
\end{eqnarray}
with which we can obtain the modular $\mathcal S$ matrix
\begin{eqnarray}
\mathcal{S} &\sim& \bar P^{(1)}\mathcal{S} \left( \bar
P^{(1)}\right)^{-1}\nonumber
\\&=&\left[P^{(2)}\mathcal{S} \left( \bar
P^{(1)}\right)^{-1}\right]^{-1}\left[P^{(2)}\mathcal{S} \left( \bar
P^{(3)}\right)^{-1}
\right]\left[P^{(3)}\mathcal{S} \left( \bar P^{(1)}\right)^{-1}\right] \nonumber\\
&=& \left\{R\left[\left(\bar U^{(2)}\right)^{-1}\bar
U^{(1)}\right]\right\}^{-1}R\left[\left(\bar U^{(2)}\right)^{-1}\bar
U^{(3)}\right]R\left[\left(\bar U^{(3)}\right)^{-1}\bar
U^{(1)}\right] \nonumber\\\label{eqn:solveS}
\end{eqnarray}
upto a trivial permutation ($=\bar P_1$ ) of the quasiparticles'
ordering sequence. A physical interpretation of Eqn. \ref{eqn:modS}
is that through our entanglement interferometry that consists of a
series of modular transformations between bases
$\vec{w}^{(1)}\rightarrow \vec{w}^{(2)}\rightarrow
\vec{w}^{(3)}\rightarrow\vec{w}^{(1)}$, one effectively cancels the
impact of undetermined quantities such as phase factors and relative
orderings.

In passing, we recall that with the help of the Verlinde's formula
\begin{eqnarray}
N^{c}_{ab}=\underset{x}{\sum}\frac{\mathcal S_{ax}\mathcal
S_{bx}\mathcal S_{\bar cx}}{\mathcal S_{1x}}
\end{eqnarray}
one can now also construct the fusion rule coefficients: $a\times b =
\underset{c}{\sum}N^{c}_{ab}$ from the obtained modular $\mathcal S$
matrix.

\section{General constraints on the modular $\mathcal U$
matrix}\label{sec:umat}

In this section, we use our three-entanglement-bipartition
construction to extract information on the modular $\mathcal{U}$
matrix, a diagonal matrix whose $a$th element encodes the
topological spin and self-statistics of the $a$ quasiparticle.
Recall\cite{zhang2012} that in the presence of the $2\pi/3$ rotation
symmetry $R_{2\pi/3}$, the modular $\mathcal{U}$ matrix is fully
determined without any ambiguity $\mathcal
{US}=\left\langle\Xi^{(1)}\right|
R_{2\pi/3}\left|\Xi^{(1)}\right\rangle$.

For simplicity, we first reorder the MESs of the second and third
entanglement bipartitions with $\bar P^{(1)}\left(\bar
P^{(2)}\right)^{-1}$ and $\bar P^{(1)}\left(\bar
P^{(3)}\right)^{-1}$, respectively, so that all $\bar P^{(\alpha)}$
are consistent and the remaining $\tilde P^{(\alpha)}$ parts only
contribute some undetermined diagonal phase factors.
Eqn.\ref{eqn:psp} is thus simplified as
\begin{eqnarray}
\mathcal
S&=&R\left[\left(\bar U^{(2)}\right)^{-1}\bar U^{(1)}\right]=\Lambda^{1L}\left(\bar U^{(2)}\right)^{-1}\bar U^{(1)}\Lambda^{1R}\nonumber\\
&=&R\left[\left(\bar U^{(3)}\right)^{-1}\bar U^{(1)}\right]=\Lambda^{2L}\left(\bar U^{(3)}\right)^{-1}\bar U^{(1)}\Lambda^{2R}\nonumber\\
&=&R\left[\left(\bar U^{(2)}\right)^{-1}\bar
U^{(3)}\right]=\Lambda^{3L}\left(\bar U^{(2)}\right)^{-1}\bar
U^{(3)} \Lambda^{3R} \label{eqn:lambda}
\end{eqnarray}
where the $\Lambda$ matrices are the diagonal phase factors used in
the function $R$ to make the elements of the first column and row of
the argument matrix real and positive. We find that the
quasiparticle topological spin $\theta_a$ is obtainable through
\begin{eqnarray}
\theta_a \propto \left[\Lambda_a^{1R}\right]^{-1} \Lambda_a^{2R}
\left( \theta_a \theta_p/\theta_{p\times
a}\right)=\left[\Lambda_a^{1R}\right]^{-1}
\Lambda_a^{2R}\left|\mathcal S_{a p}\right|/\mathcal S_{a p}
\label{eqn:modU}
\end{eqnarray}
where $\theta_{p\times a}/ \theta_a \theta_p$ is the braiding
between $a$ and some as yet undetermined Abelian quasiparticle $p$,
which is encoded as the phase factors of the elements in the $p$th
column of the modular $\mathcal S$ matrix.

To derive Eqn. \ref{eqn:modU}, we first compare Eqn. \ref{eqn:u2u1},
\ref{eqn:u3u1}, \ref{eqn:tst} and \ref{eqn:lambda} and realize that
the contributions to the diagonal phase factors originate from the
modular $\mathcal U$ matrix, the $V^{(\alpha)}$ conventions and the
remaining relative orderings $\tilde P^{(\alpha)}$. More
specifically, we have
\begin{eqnarray}
\Lambda^{1R}_{a}&\propto& V^{(1)}_a
\left(\theta_a \theta_{p_2} / \theta_{a\times p_2}\right)\nonumber \\
\Lambda^{2R}_a&\propto& V^{(1)}_a  [\tilde P^{(1)} \mathcal U
\left(\tilde P^{(1)}\right)^{-1}]_a \left(\theta_a \theta_{p_3} /
\theta_{a\times
p_3}\right)\nonumber \\
&=& V^{(1)}_a \left(\theta_{a\times p_1} \theta_a \theta_{p_3} /
\theta_{a\times p_3}\right)
\end{eqnarray}
where $p_1$, $p_2$, $p_3$ are the Abelian quasiparticles added by
$\tilde P^{(1)}$, $\tilde P^{(2)}$, $\tilde P^{(3)}$ and determined
by the particle content of the first MES of each entanglement
bipartition $\alpha = 1,2,3$, respectively. Then we can eliminate
the unknown $V^{(1)}_a$ part
\begin{eqnarray}
\left[\Lambda_a^{1R}\right]^{-1} \Lambda_a^{2R} &\propto&
\theta_{a\times p_1}\theta_{a\times p_2}/\theta_{a\times p_3}\propto
\theta_{a\times p} \label{eqn:lambda2}
\end{eqnarray}
upto some overall phases. Here $p=p_1\times p_2 \times \bar p_3$ is
also an Abelian quasiparticle where $\bar p_3$ is the anti-particle
of $p_3$.

Since the particle content of $p$ is undetermined, we need to
consider all cases where $p$ is Abelian, which gives the following
potential solutions of $\theta_a$
\begin{eqnarray}
\theta_{p\times a} \propto \theta_a (\theta_{p\times a}/ \theta_a
\theta_p)= \theta_a\mathcal S_{ap}/\left|\mathcal S_{ap}\right|
\label{eqn:uacandi}
\end{eqnarray}
together with Eqn. \ref{eqn:lambda2} we obtain our result in Eqn.
\ref{eqn:modU}. Similar expressions can be straightforwardly
obtained with $\Lambda_a^{3L} \left[\Lambda_a^{1L}\right]^{-1}$ and
$\Lambda_a^{2L} \Lambda_a^{3R}$ instead.

The overall phase of $\theta_a$ can be fixed by requiring
$\theta_1=1$ for the identity particle. In addition, more
information on $p$ can be obtained by imposing certain consistency
requirements. In particular, the self-braiding - the phase obtained
when an Abelian quasiparticle $a$ braids around another $a$ should
be twice as much as the self-statistics - the phase when two $a$
quasiparticles exchange with each other, therefore
$\theta^2_a=\theta_{a\times a}/ \theta_a \theta_a= \mathcal
S_{aa}/\left|\mathcal S_{aa}\right|$ equals the phase factor of the
$a$th diagonal element in the modular $\mathcal S$ matrix.
Therefore, if $\theta_a$ is a consistent solution, another solution
$\theta'_a=\theta_a\mathcal S_{ap}/\left|\mathcal
S_{ap}\right|=\theta_{\alpha\times a}/\theta_p$ is also consistent
if and only if $\left(\theta_{p\times
a}/\theta_p\right)^2=\left(\theta'_a\right)^2 = \mathcal
S_{aa}/\left|\mathcal S_{aa}\right|=\theta_a^2 $ for the choice of
$p$. Correspondingly, $\theta_{p\times a}/\theta_p \theta_a =
\mathcal S_{p a}/\left|\mathcal S_{p a}\right|= \pm 1$, the Abelian
elements of the $p$th column in the modular $\mathcal S$ matrix need
to be fully real. In particular, when the first column is the only
column in the modular $\mathcal S$ matrix where all Abelian elements
are real, our algorithm completely determines $\theta_a$. We show
later an example of the $\mathbb Z_3$ gauge theory where $\theta_a$
can be completely determined given the modular $\mathcal S$ matrix
and $\theta_{a\times p}$.

In addition, when the topological ordered state is bosonic, the
modular $\mathcal U$ matrix by definition $\mathcal U_a = \theta_a
\exp(-i2\pi c/24)$ and the corresponding central charge $c$ can be
determined (modulo 8) by the requirement that $\left(\mathcal
{US}\right)^3=1$.

\section{Examples: The $\mathbb Z_{2}$ gauge
theory (Toric code model), the $SU(2)_{3}$ Chern Simons theory and
the $\mathbb Z_{3}$ gauge theory}\label{sec:example}

\subsection{Obtaining the Modular $\mathcal S$ matrix of the $\mathbb Z_2$ gauge theory}

In this subsection, we use Kitaev's square lattice toric code
model\cite{kitaev_honey} as an example for our algorithm. The ground
state is an equal superposition of all possible configurations of
closed electric field loops on the lattice. On a torus, the four
degenerate ground states $|\xi_{ab}\rangle$, $a,b=0,1$ are
distinguished by the winding number parities $a$, $b$ of the
electric field loops around the two cycles of the torus and cannot
be mixed by any local operator, constituting the $\mathbb Z_2$ gauge
theory.

The nature of the MESs for the toric code model was studied in Ref.
\onlinecite{zhang2012}. For a nontrivial entanglement bipartition,
the MESs are the simultaneous eigenstates of electric and magnetic
fluxes threading the entanglement bipartition boundary. The MESs for
entanglement bipartition along the $\vec w_2^{(1)} = \hat y$
direction are
\begin{eqnarray}
|\Xi_1 \rangle = \frac{e^{i\varphi_1}}{\sqrt{2}}(|\xi_{00}\rangle + |\xi_{01}\rangle) \nonumber \\
|\Xi_2 \rangle = \frac{e^{i\varphi_2}}{\sqrt{2}}(|\xi_{00}\rangle - |\xi_{01}\rangle) \nonumber \\
|\Xi_3 \rangle = \frac{e^{i\varphi_3}}{\sqrt{2}}(|\xi_{10}\rangle + |\xi_{11}\rangle) \nonumber \\
|\Xi_4 \rangle = \frac{e^{i\varphi_4}}{\sqrt{2}}(|\xi_{10}\rangle -
|\xi_{11}\rangle) \label{eqn:mes1}
\end{eqnarray}
where $\varphi_i$ are undetermined phases for each MES. The unitary
matrix $\bar U_1$ connecting the $\vec w_2^{(1)}$ MESs and the
electric field parity states
$\left\{|\xi_{00}\rangle,|\xi_{01}\rangle,|\xi_{10}\rangle,|\xi_{11}\rangle\right\}$
is
\begin{equation}
\bar U_1  = \frac{1}{\sqrt 2} \left(
\begin{array}{cccc}
e^{i\varphi_1} & e^{i\varphi_2} &  &   \\
e^{i\varphi_1} & -e^{i\varphi_2} &  &   \\
&  & e^{i\varphi_3} & e^{i\varphi_4}  \\
&  & e^{i\varphi_3} & -e^{i\varphi_4}
\end{array} \right)
\label{eqn:u1}
\end{equation}

On the other hand, it is straightforward to verify that for
entanglement bipartition boundary along the $\vec w_2^{(2)} = -\hat
x$ direction the corresponding MESs are
\begin{eqnarray}
|\Xi_1' \rangle = \frac{e^{i\varphi_2'}}{\sqrt{2}}(|\xi_{00}\rangle - |\xi_{10}\rangle) \nonumber \\
|\Xi_2' \rangle = \frac{e^{i\varphi_3'}}{\sqrt{2}}(|\xi_{01}\rangle
+
|\xi_{11}\rangle) \nonumber \\
|\Xi_3' \rangle = \frac{e^{i\varphi_4'}}{\sqrt{2}}(|\xi_{01}\rangle
- |\xi_{11}\rangle) \nonumber \\
|\Xi_4' \rangle = \frac{e^{i\varphi_1'}}{\sqrt{2}}(|\xi_{00}\rangle
+ |\xi_{10}\rangle) \label{eqn:mes2}
\end{eqnarray}
where again $\varphi_{i}'$ are undetermined phases for each MES. We
have purposefully scrambled the ordering of the MESs so that
$|\Xi_{i} \rangle$ and $|\Xi_{i}' \rangle$ do not necessarily
correspond to the same quasiparticle and the modular $\mathcal S$
matrix is not directly obtainable from only two sets of MESs. The
unitary matrix $\bar U_2$ connecting the $\vec w_{2}^{(2)}$ MESs and
the electric field parity states is
\begin{equation}
\bar U_2  = \frac{1}{\sqrt 2} \left(
\begin{array}{cccc}
e^{i\varphi_{2}'}    &                    &                     & e^{i\varphi_{1}'}    \\
                    &e^{i\varphi_{3}'}    & e^{i\varphi_{4}'} &     \\
-e^{i\varphi_{2}'}   &                    &                    &  e^{i\varphi_{1}'}    \\
                    &e^{i\varphi_{3}'}    &   -e^{i\varphi_{4}'} &
\end{array} \right) \label{eqn:u2}
\end{equation}

Now we need to introduce just another entanglement bipartition. Let
us consider taking the boundary along the $\vec w_{2}^{(3)} = -\hat
x + \hat y$ direction so that $\vec w_2^{(3)}=\vec w_2^{(1)}+\vec
w_2^{(2)}$. The corresponding MESs are
\begin{eqnarray}
|\Xi_{1}'' \rangle = \frac{e^{i\varphi_{3}''}}{\sqrt{2}}(|\xi_{01}\rangle + |\xi_{10}\rangle) \nonumber \\
|\Xi_{2}'' \rangle =
\frac{e^{i\varphi_{4}''}}{\sqrt{2}}(|\xi_{01}\rangle -
|\xi_{10}\rangle) \nonumber \\
|\Xi_{3}'' \rangle = \frac{e^{i\varphi_{2}''}}{\sqrt{2}}(|\xi_{00}\rangle - |\xi_{11}\rangle) \nonumber \\
|\Xi_{4}'' \rangle =
\frac{e^{i\varphi_{1}''}}{\sqrt{2}}(|\xi_{00}\rangle +
|\xi_{11}\rangle) \label{eqn:mes3}
\end{eqnarray}
where $\varphi_i''$ are undetermined phases and we have once again
scrambled the ordering of the quasiparticles to make a difference
from the previous two. The unitary matrix $\bar U_3$ connecting the
$\vec w_{2}^{(3)}$ MESs and the electric field parity states is
\begin{equation}
\bar U_3  = \frac{1}{\sqrt 2} \left(
\begin{array}{cccc}
& &e^{i\varphi_{2}''}& e^{i\varphi_{1}''}   \\
e^{i\varphi_{3}''}& e^{i\varphi_{4}''} & &\\
e^{i\varphi_{3}''}& -e^{i\varphi_{4}''}& & \\
& &-e^{i\varphi_{2}''}&e^{i\varphi_{1}''}
\end{array} \right) \label{eqn:u3}
\end{equation}

From Eqn. \ref{eqn:u1}, \ref{eqn:u2} and  \ref{eqn:u3}, we can
construct matrices $\bar U_{2}^{-1}\bar U_{1}$, $\bar U_{3}^{-1}\bar
U_{1}$ and $\bar U_{2}^{-1}\bar U_{3}$. By setting the elements of
the first rows and columns to be real and positive, we find:
\begin{eqnarray}
\bar P_2^{-1}\mathcal S \bar P_1= R(\bar U_{2}^{-1}\bar U_{1})=
\frac{1}{2}\left(\begin{array}{cccc}
1 &  1 &  1 &  1\\
1 & -1 & -1 &  1\\
1 & -1 &  1 & -1\\
1 &  1 & -1 & -1\end{array} \right) \nonumber\\
\bar P_3^{-1}\mathcal S \bar P_1= R(\bar U_{3}^{-1}\bar U_{1})=
\frac{1}{2}\left(\begin{array}{cccc}
1 &  1 &  1 &  1\\
1 &  1 & -1 & -1\\
1 & -1 & -1 &  1\\
1 & -1 &  1 & -1\end{array} \right) \nonumber\\
\bar P_2^{-1}\mathcal S \bar P_3= R(\bar U_{2}^{-1}\bar U_{3})=
\frac{1}{2}\left(\begin{array}{cccc}
1 &  1 &  1 &  1\\
1 & -1 & -1 &  1\\
1 &  1 & -1 & -1\\
1 & -1 &  1 & -1\end{array} \right)
\end{eqnarray}
where $\bar P_\alpha$ are permutation matrices acting on the 2nd,
3rd and 4th columns and rows. According to Eqn. \ref{eqn:solveS}, we
obtain the following solution consistent with the $\mathbb Z_2$
gauge theory:
\begin{eqnarray}
\mathcal S= \frac{1}{2}\left(\begin{array}{cccc}
1 &  1 &  1 &  1\\
1 &  1 & -1 & -1\\
1 & -1 &  1 & -1\\
1 & -1 & -1 &  1\end{array} \right)
\end{eqnarray}
with $\bar P_1=I$ and:
\begin{eqnarray}
\bar P_2=\bar P_2^{-1}=\left(\begin{array}{cccc}
1 &    &    &   \\
  &    &    &  1\\
  &    &  1 &   \\
  &  1 &    &   \end{array} \right),
  \bar P_3=\bar P_3^{-1}=\left(\begin{array}{cccc}
1 &    &    &   \\
  &  1 &    &   \\
  &    &    &  1\\
  &    &  1 &   \end{array} \right)
\end{eqnarray}

In addition, we can obtain the diagonal $\Lambda$ matrices by
comparing the matrices $\bar U_{2}^{-1}\bar U_{1}$, $\bar
U_{3}^{-1}\bar U_{1}$ and $\bar U_{2}^{-1}\bar U_{3}$ before and
after the function $R$. In particular, we have $\Lambda^{1L}\propto
\mbox{diag}\left(e^{-i\varphi_{1}}, e^{-i\varphi_{2}},
-e^{-i\varphi_{3}}, -e^{-i\varphi_{4}}\right)$ and
$\Lambda^{2L}\propto \mbox{diag}\left(e^{-i\varphi_{1}},
-e^{-i\varphi_{2}}, e^{-i\varphi_{3}}, e^{-i\varphi_{4}}\right)$.
According to Eqn. \ref{eqn:lambda2}, this leads to:
\begin{eqnarray}
\theta_{\alpha\times a}\propto \left[\Lambda^{1R}\right]^{-1}
\Lambda^{2R}\propto \mbox{diag}\left(1, -1, -1, -1\right)
\end{eqnarray}
together with the obtained modular $\mathcal S$ matrix and Eqn.
\ref{eqn:uacandi}, the possible solutions of the quasiparticle spins
$\theta_a$ are:
\begin{eqnarray}
\theta_a&=&\left(1, 1, 1, -1\right), c=0\nonumber \\\mbox{or   }
\theta_a&=&\left(1, 1, -1,1\right), c=0\nonumber \\\mbox{or   }
\theta_a&=&\left(1, -1, 1, 1\right), c=0\nonumber \\\mbox{or   }
\theta_a&=&\left(1, -1, -1, -1\right), c=4
\end{eqnarray}
where we have used $\left(\mathcal {US}\right)^3=1$ and $\mathcal
U_a = \theta_a \exp(-i2\pi c/24)$ to extract the value of $c$. Since
all elements of the modular $\mathcal S$ matrix are real, we cannot
refine these candidates further. In Sec. \ref{sec:z3u}, we discuss
another example, the $\mathbb Z_3$ gauge theory, where there is only
one fully real column in the modular $\mathcal S$ matrix thus
$\theta_a$ can be uniquely determined.

\subsection{The non-Abelian $SU(2)_{3}$ Chern Simons theory} \label{sec:nonab}

In this subsection, we provide an example of applying our algorithm
to a non-Abelian state: the $SU(2)_{3}$ Chern Simons theory. The
modular $\mathcal{S}$ matrix of $SU(2)_{3}$ topological ordered
phase is:
\begin{equation}
\mathcal{S}=\sqrt{\frac{2}{5}}\sin\frac{\pi}{5}\left(\begin{array}{cccc}
1 & \sigma & \sigma & 1\\
\sigma & 1 & -1 & -\sigma\\
\sigma & -1 & -1 & \sigma\\
1 & -\sigma & \sigma & -1
\end{array}\right)
\end{equation}
where $\sigma=\left(1+\sqrt{5}\right)/2$ is the golden ratio.
Corresponding to each column (row), the four quasiparticles have
quantum dimensions $d=1,\sigma,\sigma,1$ respectively -- clearly the
second and third quasiparticles are non-Abelian. If we introduce
only two entanglement bipartitions, the TEE can make a distinction
between the MESs associated with the Abelian quasiparticles and
those with the non-Abelian quasiparticles, yet it can not
distinguish between the first and fourth MESs, both associated with
Abelian quasiparticles, as well as between the second and third
MESs, both associated with non-Abelian quasiparticles of equal
quantum dimensions.

Following the algorithm in the main text, we introduce three
entanglement bipartitions along the $\vec{w}^{(1)}$, $\vec{w}^{(2)}$
and $\vec{w}^{(3)}=\vec{w}^{(1)}+\vec{w}^{(2)}$ directions, and
shuffle their respective ordering of the MESs with the only
requirement that the first MES is either
$\left|\Xi_{1}^{(i)}\right\rangle $ or
$\left|\Xi_{4}^{(i)}\right\rangle $, which are associated with
Abelian quasiparticles. As a particular example: MESs along the
$w^{(1)}$: $\left\{ \left|\Xi_{4}^{(1)}\right\rangle
,\left|\Xi_{3}^{(1)}\right\rangle ,\left|\Xi_{2}^{(1)}\right\rangle
,\left|\Xi_{1}^{(1)}\right\rangle \right\} $ ; MESs along the
$w^{(2)}$: $\left\{ \left|\Xi_{4}^{(2)}\right\rangle
,\left|\Xi_{2}^{(2)}\right\rangle ,\left|\Xi_{3}^{(2)}\right\rangle
,\left|\Xi_{1}^{(2)}\right\rangle \right\} $ ; MESs along the
$w^{(3)}$: $\left\{ \left|\Xi_{1}^{(3)}\right\rangle
,\left|\Xi_{3}^{(3)}\right\rangle ,\left|\Xi_{2}^{(3)}\right\rangle
,\left|\Xi_{4}^{(3)}\right\rangle \right\} $ . After neutralizing
all the diagonal phase factors in Eq. 6 and 10 from the modular
$\mathcal{U}$ matrix and the $V^{(i)}$ conventions, we obtain the
transformation between these MES bases:
\begin{eqnarray}
R\left[\left(\bar{U}^{(3)}\right)^{-1}\bar{U}^{(1)}\right] & = &
\sqrt{\frac{2}{5}}\sin\frac{\pi}{5}R\left[\left(\begin{array}{cccc}
1 & \sigma & \sigma & 1\\
\sigma & -1 & -1 & \sigma\\
-\sigma & -1 & 1 & \sigma\\
-1 & \sigma & -\sigma & 1
\end{array}\right)\right] \nonumber\\&= &\sqrt{\frac{2}{5}}\sin\frac{\pi}{5}\left(\begin{array}{cccc}
1 & \sigma & \sigma & 1\\
\sigma & -1 & -1 & \sigma\\
\sigma & 1 & -1 & -\sigma\\
1 & -\sigma & \sigma & -1
\end{array}\right)\nonumber\\
R\left[\left(\bar{U}^{(2)}\right)^{-1}\bar{U}^{(3)}\right]& =
&\sqrt{\frac{2}{5}}\sin\frac{\pi}{5}R\left[\left(\begin{array}{cccc}
1 & \sigma & -\sigma & -1\\
\sigma & -1 & 1 & -\sigma\\
\sigma & -1 & -1 & \sigma\\
1 & \sigma & \sigma & 1
\end{array}\right)\right] \nonumber \\ & = &\sqrt{\frac{2}{5}}\sin\frac{\pi}{5}\left(\begin{array}{cccc}
1 & \sigma & \sigma & 1\\
\sigma & -1 & -1 & \sigma\\
\sigma & -1 & 1 & -\sigma\\
1 & \sigma & -\sigma & -1
\end{array}\right)\nonumber\\
R\left[\left(\bar{U}^{(2)}\right)^{-1}\bar{U}^{(1)}\right] &= &
\sqrt{\frac{2}{5}}\sin\frac{\pi}{5}R\left[\left(\begin{array}{cccc}
-1 & \sigma & -\sigma & 1\\
-\sigma & -1 & 1 & \sigma\\
\sigma & -1 & -1 & \sigma\\
1 & \sigma & \sigma & 1
\end{array}\right)\right]\nonumber \\ & = & \sqrt{\frac{2}{5}}\sin\frac{\pi}{5}\left(\begin{array}{cccc}
1 & \sigma & \sigma & 1\\
\sigma & -1 & -1 & \sigma\\
\sigma & 1 & -1 & -\sigma\\
1 & -\sigma & \sigma & -1
\end{array}\right)
\end{eqnarray}

Then, it is straightforward to check Eq. 7 gives the consistent
modular $\mathcal{S}$ matrix:
\begin{eqnarray}
& & \left\{
R\left[\left(\bar{U}^{(2)}\right)^{-1}\bar{U}^{(1)}\right]\right\}
^{-1}R\left[\left(\bar{U}^{(2)}\right)^{-1}\bar{U}^{(3)}\right]R\left[\left(\bar{U}^{(3)}\right)^{-1}\bar{U}^{(1)}\right] \nonumber  =  \\
& & \sqrt{\frac{2}{5}}\sin\frac{\pi}{5}\left(\begin{array}{cccc}
1 & \sigma & \sigma & 1\\
\sigma & 1 & -1 & -\sigma\\
\sigma & -1 & -1 & \sigma\\
1 & -\sigma & \sigma & -1
\end{array}\right)
\end{eqnarray}

\subsection{Obtaining the modular $\mathcal U$ matrix of the $\mathbb Z_3$ gauge
theory}\label{sec:z3u}

In this subsection, we briefly introduce another example where the
modular $\mathcal S$ matrix together with $\theta_{a\times \alpha}$
completely determines the values of $\theta_a$ even if the Abelian
quasiparticle $\alpha$ is yet undetermined.

The modular $\mathcal S$ matrix of the Abelian $\mathbb Z_3$ gauge
theory, which is fully obtainable by similar argument to the last
section, is:
\begin{eqnarray}
\mathcal{S}=\frac{1}{3}\left(\begin{array}{ccccccccc}
1 & 1 & 1 & 1 & 1 & 1 & 1 & 1 & 1\\
1 & 1 & 1 & q & q & q & q^{2} & q^{2} & q^{2}\\
1 & 1 & 1 & q^{2} & q^{2} & q^{2} & q & q & q\\
1 & q & q^{2} & 1 & q & q^{2} & 1 & q & q^{2}\\
1 & q & q^{2} & q & q^{2} & 1 & q^{2} & 1 & q\\
1 & q & q^{2} & q^{2} & 1 & q & q & q^{2} & 1\\
1 & q^{2} & q & 1 & q^{2} & q & 1 & q^{2} & q\\
1 & q^{2} & q & q & 1 & q^{2} & q^{2} & q & 1\\
1 & q^{2} & q & q^{2} & q & 1 & q & 1 & q^{2}
\end{array}\right)
\end{eqnarray}
where $q=e^{i2\pi/3}$. Its diagonal elements as the self-braiding
should be consistent with the self-statistics of the quasiparticles
with presumed ordering of $(1,e,e^2,m,em,e^2m,m^2,em^2,e^2m^2)$:
\begin{eqnarray}
\theta^2_{a}=(1, 1, 1, 1, q^2, q, 1, q, q^2) \label{eqn:z3s}
\end{eqnarray}

On the other hand, given $\theta_{a\times p}$ without knowing the
actual Abelian particle content of $p$, we can derive the following
possibilities for $\theta_a$ according to Eqn. \ref{eqn:uacandi} by
considering each column of the modular $\mathcal S$ matrix above:
\begin{eqnarray}
\theta_{a}&=&(1, 1, 1, 1, q, q^2, 1, q^2, q) \nonumber
\\\mbox{or   }
\theta_{a}&=&(1, 1, 1, q, q^2, 1, q^2, q, 1) \nonumber
\\\mbox{or   }
\theta_{a}&=&(1, 1, 1, q^2, 1, q, q, 1, q^2) \nonumber
\\\mbox{or   }
\theta_{a}&=&(1, q, q^2, 1, q^2, q, 1, 1, 1) \nonumber
\\\mbox{or   }
\theta_{a}&=&(1, q, q^2, q, 1, q^2, q^2, q^2, q^2) \nonumber
\\\mbox{or   }
\theta_{a}&=&(1, q, q^2, q^2, q, 1, q, q, q) \nonumber
\\\mbox{or   }
\theta_{a}&=&(1, q^2, q, 1, 1, 1, 1, q, q^2) \nonumber
\\\mbox{or   }
\theta_{a}&=&(1, q^2, q, q, q, q, q^2, 1, q) \nonumber
\\\mbox{or   }
\theta_{a}&=&(1, q^2, q, q^2, q^2, q^2, q, q^2, 1)
\end{eqnarray}
however, only the first candidate is consistent with Eqn.
\ref{eqn:z3s} from the modular $\mathcal S$ matrix. Therefore the
$\theta_a$ solution is unique, and the statistics of the
quasiparticles can be uniquely determined. In addition, with
$\left(\mathcal{US}\right)^3=1$ we can obtain the modular $\mathcal
U$ matrix: $\mathcal U=\mbox{diag}(\theta_a)=(1, 1, 1, 1, q, q^2, 1,
q^2, q)$.

\section{Conclusion}

In this paper, we extended the discussion in Ref.\cite{zhang2012} to
characterize a two-dimensional topological ordered phase with only
its complete set of ground-state wavefunctions. Based on a closed
sequence of modular transformations between three inequivalent
entanglement bipartitions, our algorithm derives the modular
$\mathcal S$ matrix and the corresponding quasiparticle braiding of
topologically ordered phase without presuming any lattice
symmetries. It also constrains the modular $\mathcal U$ matrix to a
few discrete possibilities, and in certain cases determines it
fully. Our algorithm is applicable to Abelian and non-Abelian phases
alike.

In general, however, our algorithm still does not guarantee a
definitive solution to the modular $\mathcal U$ matrix and thus, the
quasiparticle self-statistics. For chiral phases, momentum
polarization can determine the self-statistics and the chiral edge
central charge\cite{tu2013}. Recently, another method to obtain the
modular matrices based on universal wavefunction
overlap\cite{wen2014} has been introduced, yet limited to small
system sizes due to an exponentially small prefactor. A universally
applicable method for the modular $\mathcal U$ matrix is still under
study.

We would like to thank Vedika Khemani for penetrating questions
regarding earlier work and Xiao-Liang Qi, Jeongwan Haah and Andre
Broido for insightful discussions. This work is supported by the
Stanford Institute for Theoretical Physics(YZ), the National Science
Foundation under Grant No. NSF PHY11-25915(TG) and NSF-DMR
1206728(AV). T.G. is supported by Gordon and Betty Moore Foundation
fellowship under the EPiQS initiative.

\appendix

\section{Results for general entanglement bipartitions} \label{sec:generalbipart}

In this Appendix, we discuss the generalization to our choices of
entanglement bipartitions in Eqn. \ref{eqn:w1w2}, \ref{eqn:w2w3} and
\ref{eqn:w1w3} as well as that of Fig. \ref{fig1} and \ref{fig2}.
Without loss of generality, we can always define:
\begin{eqnarray}
\vec{w}_{1}^{(3)}&=&n_{1}\vec{w}_{1}^{(1)}+m_{1}\vec{w}_{2}^{(1)}
\nonumber \\
\vec{w}_{2}^{(3)}&=&n_{2}\vec{w}_{1}^{(1)}+m_{2}\vec{w}_{2}^{(1)}
\label{eqn:oldw1w2}
\end{eqnarray}
and:
\begin{eqnarray}
\vec{w}_{1}^{(2)}=n_{3}\vec{w}_{1}^{(3)}+m_{3}\vec{w}_{2}^{(3)}
\nonumber \\
\vec{w}_{2}^{(2)}=n_{4}\vec{w}_{1}^{(3)}+m_{4}\vec{w}_{2}^{(3)}
\label{eqn:oldw2w3}
\end{eqnarray}
with $n_{1}m_{2}-m_{1}n_{2}=1$ and $n_{3}m_{4}-m_{3}n_{4}=1$ by
definition of the modular transformation.

First, in order to make our algorithm work, we would like all these
transformations between different MES bases to be the form of the
modular $\mathcal S$ matrix times some diagonal phase factors, which
requires the cross products
$\vec{w}_{2}^{(1)}\times\vec{w}_{2}^{(2)}=\vec{w}_{2}^{(1)}\times\vec{w}_{2}^{(3)}=\vec{w}_{2}^{(3)}\times\vec{w}_{2}^{(2)}=A$
where $A$ is the (signed) surface area of the torus. For example,
for the transformation in Eqn. \ref{eqn:oldw1w2} this requires
$n_{2}=-1$. Then the corresponding $\ensuremath{SL(2,\mathbb{Z})}$
matrix has the following expansion:
\begin{eqnarray}F=\left(\begin{array}{cc}
n_{1} & 1-n_{1}m_{2}\\
-1 & m_{2}
\end{array}\right)&=&\left(\begin{array}{cc}
1 & -n_{1}\\
 & 1
\end{array}\right)\left(\begin{array}{cc}
 & 1\\
-1
\end{array}\right)\left(\begin{array}{cc}
1 & -m_{2}\\
 & 1
\end{array}\right) \nonumber\\
&=&U^{-n_1}SU^{-m_2} \label{eqn:sl12}\end{eqnarray} thus the modular
matrix is $\mathcal
F_{13}\mathcal{(S,U)}=\mathcal{U}^{-n_1}\mathcal{S}\mathcal{U}^{-m_2}$.

Similarly we require $n_{4}=-1$. It is also straightforward to
derive the $\ensuremath{SL(2,\mathbb{Z})}$ matrix for the
transformation from the $\vec{w}^{(1)}$ to $\vec{w}^{(2)}$ from the
product of Eqn. \ref{eqn:oldw1w2} and Eqn. \ref{eqn:oldw2w3}:
\begin{eqnarray}& &\left(\begin{array}{cc}
n_{3} & 1-n_{3}m_{4}\\
-1 & m_{4}
\end{array}\right)\left(\begin{array}{cc}
n_{1} & 1-n_{1}m_{2}\\
-1 & m_{2}
\end{array}\right)\nonumber\\& &=\left(\begin{array}{cc}
n_{3}n_1+n_3 m_4-1 & n_3-n_1 n_3 m_2 + m_2 - n_3 m_2 m_4\\
-n_1-m_4 & m_2 m_{4}+n_1 m_2 -1
\end{array}\right) \end{eqnarray}
therefore we have another requirement $-n_{1}-m_{4}=-1$.

Under these constraints, the $\ensuremath{SL(2,\mathbb{Z})}$
matrices for the transformation from the $\vec{w}^{(3)}$ to
$\vec{w}^{(2)}$ and $\vec{w}^{(1)}$ to $\vec{w}^{(2)}$ may be
expanded in terms of generators in Eqn. \ref{eqn:sl2gen} as:
\begin{eqnarray}
\left(\begin{array}{cc}
n_{3} & 1-n_{3}+n_{1}n_{3}\\
-1 & 1-n_{1}
\end{array}\right)=\left(\begin{array}{cc}
1 & -n_{3}\\
 & 1
\end{array}\right)\left(\begin{array}{cc}
 & 1\\
-1
\end{array}\right)\left(\begin{array}{cc}
1 & n_{1}-1\\
 & 1
\end{array}\right)\label{eqn:sl23}\end{eqnarray}
and:
\begin{eqnarray}
\left(\begin{array}{cc}
n_{3}-1 & n_{3}+m_{2}-n_{3}m_{2}\\
-1 & m_{2}-1
\end{array}\right)=\left(\begin{array}{cc}
1 & 1-n_{3}\\
 & 1
\end{array}\right)\left(\begin{array}{cc}
 & 1\\
-1
\end{array}\right)\left(\begin{array}{cc}
1 & 1-m_{2}\\
 & 1
\end{array}\right)\nonumber \\ \label{eqn:sl13}\end{eqnarray}

The corresponding modular matrices for the transformations between
the sets of MESs are:
\begin{eqnarray}
\mathcal
F_{23}\mathcal{(S,U)}=\mathcal{U}^{-n_3}\mathcal{S}\mathcal{U}^{n_1-1}
\label{eqn:oldu3u2}
\end{eqnarray}
and
\begin{eqnarray}
\mathcal
F_{12}\mathcal{(S,U)}=\mathcal{U}^{1-n_3}\mathcal{S}\mathcal{U}^{1-m_2}
\label{eqn:oldu3u1}
\end{eqnarray}
respectively.

Especially, it is clear that
$\vec{w}_{2}^{(3)}=-\vec{w}_{1}^{(1)}+m_{2}\vec{w}_{2}^{(1)}$ and
$\vec{w}_{2}^{(2)}=-\vec{w}_{1}^{(1)}+(m_{2}-1)\vec{w}_{2}^{(1)}$,
thus we have to have
$\vec{w}_{2}^{(3)}=\vec{w}_{2}^{(1)}+\vec{w}_{2}^{(2)}$. This
explains our choice of the third entanglement bipartition in the
main text.

The remaining degrees of freedom $n_1$, $m_2$ and $n_3$ are
concerned only with the choices of the $\vec w^{(\alpha)}_1$
directions and bring no essential change to our line of reasoning.
The specific choice in the main text corresponds to $n_1=m_2=n_3=1$.

\end{document}